\documentclass[a4paper,10pt]{amsart}

\usepackage{cite}
\usepackage{tocvsec2}
\usepackage[usenames,dvipsnames]{xcolor}

\usepackage[colorlinks, 
            bookmarks=true,bookmarksopen=true]{hyperref}
          \hypersetup
                {
                    colorlinks=true,
                    linkcolor=MidnightBlue,
                    urlcolor=MidnightBlue,
                    filecolor=black,
                    citecolor=RedOrange,
                    pdfstartview=FitV,
                    pdftitle={},
                    pdfauthor={Ilmar Gahramanov},
                    pdfsubject={},
                    pdfkeywords={},
                    pdfpagemode=None,
                    bookmarksopen=true
                }

\theoremstyle{definition}

\theoremstyle{remark}

\numberwithin{equation}{section}

\begin{document}

\rightline{\vbox{\small\hbox{\tt HU-EP-15-14} }}
 \vskip 2.2 cm

\title{Mathematical structures behind supersymmetric dualities}

\author{Ilmar Gahramanov}
\address{Particle Physics Research Group, Institute of Radiation Problems, Azerbaijan National Academy of Sciences, B.Vahabzade 9, AZ1143 Baku, Azerbaijan}
\address{Institut f\"{u}r Physik und IRIS Adlershof, Humboldt-Universit\"{a}t zu Berlin, Zum Grossen Windkanal 6, D12489 Berlin, Germany}

\email{ilmar@physik.hu-berlin.de}




\begin{abstract}
The purpose of these notes is to give a short survey of an interesting connection between partition functions of supersymmetric gauge theories and hypergeometric functions and to present the recent progress in this direction. 
\end{abstract}

\maketitle

\section{Introduction}

In these notes, we discuss several properties of basic, hyperbolic and elliptic hypergeometric functions, in particular some interesting integral identities satisfied by these functions and their relation to supersymmetric dualities in three and four dimensions. 

We will mainly focus on elliptic hypergeometric functions. The theory of these functions is quite a new research area in mathematics. The first example of the elliptic analogues of hypergeometric series was discovered about 20 years ago by Frenkel and Turaev \cite{frenkel1997elliptic} in the context of elliptic 6j-symbol. This family of functions is the top level of hypergeometric functions. Recently they have attracted the attention of physicists since they proved to be useful tool in theoretical and mathematical physics. They appear in various ways in physics, in particular the elliptic hypergeometric integrals associated to root systems\footnote{ Note that in recent years considerable progress has been made in the subject of elliptic hypergeometric functions associated to root systems, see e.g. \cite{rosengren2004,Rosengren11, rains2003,BultW4,Spiridonovbeta,SpiridonovV, Spiridonovessay}. } arise naturally in the computation of the so--called superconformal indices of four--dimensional supersymmetric gauge theories \cite{Spiridonov:2008zr,Spiridonov:2009za,Spiridonov:2010hh,Spiridonov:2011hf,Gahramanov:gka,Gahramanov:2013xsa}.

The superconformal index was introduced by R\"{o}melsberger \cite{Romelsberger:2005eg} and independently by Kinney et al. \cite{Kinney:2005ej} in 2005 as a non--trivial generalization of the Witten index\footnote{The usual Witten index measures the difference between number of fermionic and bosonic states \cite{Witten:1982df}.}. Soon later, Dolan and Osborn \cite{Dolan:2008qi} recognized  that the superconformal index can be expressed in terms of elliptic hypergeometric integral. Equalities of superconformal indices of supersymmetric dual theories lead to various complicated integral identities for the elliptic hypergeometric integrals. Some of them were known earlier, but most of them are not yet proven. 

There is a similar story for three-dimensional supersymmetric gauge theories. Namely three-dimensional superconformal index can be expressed in terms of basic hypergeometric integrals and three-dimensional sphere partition function has a form of hyperbolic hypergeometric integral (see, e.g. \cite{Gahramanov:gka, Gahramanov:2013rda,Krattenthaler:2011da,Gahramanov:2014ona,Kapustin:2011jm, Amariti:2014iza}).

The present notes are not a comprehensive review. Each section covers only few aspects of applications of basic and elliptic hypergeometric functions in physics. Here we mainly focus on non-trivial integral identities coming from supersymmetric dualitities. 

The rest of the paper is organized in the following way. In Section 2 and 3 we make a brief review of hypergeometric functions and the superconformal index, respectively. We present some
examples of non-trivial integral identities in Sections 4-6.

\section{What is an elliptic hypergeometric function?}

In this section we recall a definition of hypergeometric function. A good reference for this introductory section is the book \cite{gasper} by Gasper and Rahman and a review article \cite{Spiridonovessay} by Spiridonov. 

Let $c_n$ be complex numbers. Consider a formal power series\footnote{We call it ``formal'' since we are not interested in the convergence of the series.}
\begin{equation} \label{formalseries}
\sum_{n=0}^{\infty} c_n x^n \;.
\end{equation}
Depending on the following ratio
\begin{equation} \label{ratio}
\frac{c_{n+1}}{c_n}
\end{equation}
we define three family of hypergeometric functions. 

\textbf{Definition.} The series (\ref{formalseries}) is called
\begin{itemize}
\item an ordinary hypergeometric series if (\ref{ratio}) is a rational function of $n$; 
\item a basic hypergeometric (or simply q-hypergeometric) series if (\ref{ratio}) is a trigonometric function of $n$; 
\item an elliptic hypergeometric series if (\ref{ratio}) is an elliptic  function of $n$.
\end{itemize}
The integral representations of hypergeometric functions can be defined similarly. For instance, a contour integral $\int_{C} \Delta(u) du$ is called elliptic hypergeometric integral\footnote{Similarly one can make a definition for multivariative case.} if the meromorphic kernel $\Delta(u)$ is the solution of the following first order finite difference equation
\begin{equation}
\Delta(u+a) = h(u;b,c) \Delta(u) \;,
\end{equation}
where $a \in \mathbb{C}$ and $h(u;b,c)$ is an elliptic function with periods $b,c \in \mathbb{C}$ and $\text{Im}(b/c)\neq 0$. 

The first two types of hypergeometric functions have been  known for a long time, so we are not going to discuss them in detail. We only want to mention that in the context of supersymmetric theories ordinary hypergeometric functions appear as sphere partition function of two-dimensional supersymmetric theories \cite{Doroud:2012xw, Benini:2012ui}. Basic hypergeometric functions have proved useful in many branches of physics and we will see their role in supersymmetric theories in Section 5.

To give an example of an elliptic hypergeometric integral, let us consider the elliptic beta integral. First we need to introduce the so-called elliptic gamma function since it is convenient to express in terms of it the general class of elliptic hypergeometric integrals. The elliptic gamma function\footnote{For generalizations of this function, see \cite{nishizawa2001elliptic, narukawa2004modular, Tizzano:2014roa}. } is the meromorphic function of three complex variables
\begin{equation}
\Gamma(z,p,q) = \prod_{i,j=0}^{\infty} \frac{1-z^{-1} p^{i+1} q^{j+1}}{1-z p^i q^j}, \; \; ~~~~ \text{with $|p|,|q|<1$.}
\end{equation} 
introduced in \cite{Ruijsenaars:1997:FOA} and studied in great detail in \cite{Felder200044}. We also recall the q-Pochhammer symbol, defined as  the infinite product
\begin{equation}
(z,q)_{\infty} = \prod_{i=0}^{\infty} (1- z q^i) \;.
\end{equation}

Now we can write the simplest identity for elliptic hypergeometric integrals. Spiridonov \cite{Spiridonovbeta} has evaluated the following integral as an elliptic analog of the Euler beta integral\footnote{There is a vast literature on q-beta integrals. The
interested reader is referred to \cite{al-salam1988,Askeybeta}.}.

\textbf{Theorem} (Spiridonov). Let $t_1, \dots ,t_6,p,q \in {\mathbb{C}}$ with $|t_1|, \dots , |t_6|,|p|,|q| <1$. Then 
\begin{equation} \label{betaint}
\frac{(p;p)_\infty (q;q)_\infty}{2} \int_{\mathbb{T}} \frac{\prod_{i=1}^6 \Gamma(t_i z ;p,q)\Gamma(t_i z^{-1} ;p,q)}{\Gamma(z^{2};p,q) \Gamma(z^{-2};p,q)} \frac{dz}{2 \pi \textup{i} z} = \prod_{1 \leq i < j \leq 6} \Gamma(t_i t_j;p,q),
\end{equation}
where the unit circle $\mathbb{T}$ is taken in the positive orientation and we imposed the balancing condition $\prod_{i=1}^6 t_i=pq$. 

Limits of the elliptic beta integral lead to many identities for hypergeometric integrals\footnote{For other limits of elliptic hypergeometric functions, see, e.g. \cite{BR1,BR2,BR3}.}. For instance, if we take the limit $p \rightarrow 0$ then (\ref{betaint}) reduces to the  Nassrallah--Rahman trigonometric beta integral \cite{Nasrallah}\footnote{Note that the integral identity presented here was observed by Rahman in \cite{Rahman2} as a special case of the integral found in \cite{Nasrallah}. This integral is an extension of the well-known Askey--Wilson integral \cite{askey1985some}. If we let the $q$ tend to $1$ one obtains the corresponding ordinary hypergeometric function.}
\begin{equation}
\frac{(q,q)_\infty}{2} \int_{\mathbb{T}}\frac{(z \prod_{i=1}^5 t_i,q)_\infty (z^{-1} \prod_{i=1}^5 t_i,q)_\infty (z^2,q)_\infty (z^{-2},q)_\infty}{\prod_{i=1}^5 (t_i z)_\infty (t_i z^{-1})_\infty} \frac{dz}{2\pi i z} \ = \ \frac{\prod_{j=1}^5 (\frac{t_1 t_2 t_3 t_4 t_5}{t_j},q)_\infty}{\prod_{1 \leq i < j \leq 5} (t_i t_j,q)_\infty}
\end{equation} 
We will come back to the physical interpretation of the presented elliptic beta integral in the next sections.

\section{The superconformal index}

In this section we summarize the relevant background material on the superconformal index for theories with four supercharges in four-dimensional (${\mathcal N}=1$) and  in three-dimensional (${\mathcal N}=2$)\footnote{Here ${\mathcal N}$ denotes the amount of supersymmetry. There are many interesting theories with extended supersymmetry, however discussion of those theories is beyond the aim of the present work.}. The references for this introductory section are \cite{Spiridonov:2009za,Krattenthaler:2011da,Gahramanov:2014ona,Gahramanov:gka, Yamazaki:2013fva}.

For the benefit of readers unfamiliar with the field let us briefly summarize the basic ingredients which they need to know about supersymmetric gauge theories with four supercharges. These theories have a gauge group $G$ and a global symmetry group $F$. The gauge group multiplets belong to the adjoint representation of $G$ whereas matter multiplets belong to a suitable representation of $G$ and $F$. The supersymmetry algebra contains the $U(1)$ R-symmetry  which is in the same supermultiplet as the stress-energy tensor. We choose $R$-charges to be $r$ for all fields and $1$ for gauge fields. 

Now let us consider a generic four-dimensional ${\mathcal N}=1$ superconformal theory on $S^3 \times S^1$. In presence of the conformal symmetry the number of supercharges is doubled and the theory features supercharges $Q_{\alpha}, \bar{Q}_{\dot{\alpha}}$ and superconformal charges $S_{\alpha}, \bar{S}_{\dot{\alpha}}$, where $\alpha, \dot{\alpha}=1,2$ denotes the spins $SU(2)_1 \times SU(2)_2$ of the isometry of $S^3$. To define the superconformal index we pick one of the supercharges, say $Q=\bar{Q}_1$ and its conjugate $Q^{\dagger}=-\bar{S}_1$, which satisfies the following relation\footnote{For the full algebra, see e.g. \cite{Spiridonov:2009za}.}
\begin{equation}
\{Q,Q^{\dagger}\} = E-2\bar{J}_3-\frac32 R
\end{equation}
The four-dimensional ${\mathcal N}=1$ superconformal index is defined\footnote{Note that one can define the twisted partition function on $S^3 \times S^1$ for any supersymmetric theory. In case that theory flows to a superconformal field theory in infrared regime, the partition function computes the superconformal index. Since we define the index for a radially quantized theory one can use the so-called supersymmetric localization technique \cite{Pestun:2007rz} to compute it.} as \cite{Romelsberger:2005eg}
\begin{equation} \label{Ind}
I(\{t_i\},p,q) \ = \text{Tr}\left[ (-1)^F e^{-\beta\{Q,Q^\dagger\} }p^{R/2+\bar{J}_3+J_3} q^{R/2+\bar{J}_3-J_3} \prod t_i^{F_i} \right]
\end{equation}
where $p$ and $q$ are the complex fugacities, $F_i$ are the Cartan generators of the global symmetry group\footnote{ All the generators $F_i$ of global charges  commute with $Q$ and $Q^{\dagger}$.}, and $t_i$ are additional regulators (fugacities) corresponding to the global symmetry. The trace is taken over the full Hilbert space of the theory on $S^3$, however, only states obeying $\{Q,Q^\dagger\}=0$ contribute to the index.

As a rule, in the definition of the superconformal index by global symmetries one means the continuous symmetries of the theory. In principle, one can define such an index which respects also discrete symmetries, see e.g. \cite{Zwiebel:2011wa}.  

The superconformal index can be computed in the free field limit by using representations of the superconformal algebra and group-theoretical data of a theory \cite{Romelsberger:2005eg, Romelsberger:2007ec}. Dolan and Osborn discovered \cite{Dolan:2008qi} that the superconformal index of four-dimensional ${\mathcal N}=1$ theory can be expressed in terms of elliptic hypergeometric integrals. We will come back to this observation in the next section. 

Next we consider the superconformal index of three-dimensional ${\mathcal N}=2$ superconformal field theories.  The superconformal index in this case is defined \cite{Bhattacharya:2008bja,Kim:2009wb,Krattenthaler:2011da} in a similar way to (\ref{Ind}) as
\begin{equation}
I (\{t_i\}, q) \ = \ \text{Tr} \left[ (-1)^F e^{-\beta \{Q,Q^\dagger\}} q^{\frac12 (\Delta+j_3)} \prod_{i=1}t_i^{F_i}  \right] \;,
\end{equation}
where $\{Q,Q^\dagger\} = \Delta-R-j_3$, $\Delta$,$j_3$ and $R$ are the energy, the third component of the angular momentum on two-sphere and the $R$-charge, respectively. The fugacities $t_i$ are associated with the flavor group. 

In the case of three dimensions the superconformal index has the form of a basic hypergeometric function. We will discuss an example in Section 5.

\section{Seiberg duality via elliptic hypergeometric functions}

The superconformal index technique provides the most rigorous mathematical check of various supersymmetric dualities in various dimensions and it is the main tool for establishing new dualities as well. 

In the 1990's Seiberg \cite{Seiberg:1994pq} and many others found a non-trivial quantum equivalence between different supersymmetric theories, called supersymmetric duality \footnote{Widely known as Seiberg duality.}. To be more precise it was shown that two (or more) different theories may describe the same physics in their infrared fixed points. The identification of superconformal indices of such dual theories gives highly non-trivial integral identities for  elliptic hypergeometric functions. 

As an example, let us consider the initial Seiberg duality for supersymmetric quantum chromodynamics \cite{Dolan:2008qi,Spiridonov:2008zr,Gahramanov:2013xsa}. The following two theories flow to the same limit in the infrared asymptotics:
\begin{itemize}
\item \textbf{Theory A}: with SU(2) gauge group and quark superfields in the fundamental representation of the SU(6) flavor group. This theory has the following superconformal index
\begin{equation} \label{theoryA}
\frac{(p;p)_{\infty} (q;q)_{\infty}}{2} \int_{{\mathbb T}}
\frac{\prod_{j=1}^6 \Gamma( (pq)^{\frac16}t_j z;p,q) \Gamma((pq)^{\frac16}t_j z^{-1};p,q)}{\Gamma(z^{ 2};p,q) \Gamma(z^{-2};p,q)}
\frac{dz}{2 \pi i z}
\end{equation}
where the numerator is the contribution of chiral multiplets and the denominator is the contribution of a vector multiplet. The integration over the gauge group picks up gauge-invariant states. 	
	
	\vspace{0.3cm}
	
	\item \textbf{Theory B}: with the same flavor group and without gauge degrees of freedom, the matter sector contains meson supermultiplets in 15-dimensional antisymmetric SU(6)-tensor representation of the second rank. The index of this theory is
\begin{equation} \label{theoryB}
	\prod_{1 \leq i < j \leq 6}\Gamma(t_i t_j;p,q) \;.
\end{equation}

\end{itemize}

Since the theories described above are equivalent in their infrared conformal fixed points, their superconformal indices must match. In fact the identity (\ref{betaint}) shows the identification of the superconformal indices (\ref{theoryA}) and (\ref{theoryB}). In general, the identification of superconformal indices of dual theories in four dimensions is nothing but the Weyl symmetry transformations for certain elliptic hypergeometric functions. 

Now we want to point out some other key features of the superconformal index using the properties of elliptic hypergeometric functions. 

The superconformal index of a theory with a flavor group $F$ has the  Weyl group symmetry $W(F)$. The Weyl symmetry of the flavor group refers to the symmetry with respect to the exchange of the flavors defined in the suitable representation of the flavor group. In cases when the theory has a hidden symmetry, the coefficients in the decomposition of the superconformal index into characters of the flavor group give the sums of dimensions of irreducible representations of the larger symmetry group. One can use this property to study global symmetry enhancement in supersymmetric gauge theories.

In our example the superconformal index (\ref{theoryA}) has the Weyl group of the exceptional root system $E_6$. It means that the theory with flavor group $SU(6)$ can be extended to $E_6$ symmetry. Indeed this is a manifestation of the four-dimensional boundary model coupled to the free five-dimensional hypermultiplet with the enhanced $E_6$ flavor symmetry \cite{Gahramanov:2013xsa}\footnote{This work was inspired by the paper \cite{Dimofte:2012pd} where a similar analysis was performed for the theory with 4 flavors (see also \cite{Spiridonov:2012de}).}
\begin{align} \label{4d/5d}
I_{4d/5d} &=\prod_{1\leq i<j \leq 6} \frac{1}{\left((p q)^{\frac23}(t_i t_j)^{-1};p,q\right)_{\infty}} \prod_{i=1}^6 \frac{1}{\left((p q)^{\frac13} t_i^{-1} w^{\pm 1};p,q \right)_{\infty}} \nonumber \\  
& \quad \times \frac{(p,p)_\infty (q,q)_\infty}{2} \oint \frac{dz}{2 \pi i z} \frac{\prod_{i=1}^6 \Gamma(\sqrt[6]{p q}t_i z;p,q) \Gamma(\sqrt[6]{p q}t_i z^{-1};p,q)}{\Gamma(z^{ 2};p,q)\Gamma(z^{- 2};p,q)} \; .
\end{align}
where we introduced the shorthand notation
\begin{equation}
(z;p,q) \ : = \ \prod_{i,j=0} (1-z p^i q^j).
\end{equation}
In the expression (\ref{4d/5d}) the term 
\begin{equation}
\prod_{1\leq i<j \leq 6} \frac{1}{\left((p q)^{\frac23}(t_i t_j)^{-1};p,q\right)_{\infty}} \prod_{i=1}^6 \frac{1}{\left((p q)^{\frac13} t_i^{-1} w^{\pm 1};p,q \right)_{\infty}}
\end{equation}
corresponds to the contribution of the five-dimensional hypermultiplet. By setting all flavor fugacities to 1 and redefining $p=t^3 y$, $q=t^3 y^{-1}$ one can easily read off the $E_6$ symmetry of the superconformal index 
\begin{equation}
I_{4d/5d}=1+27 t^2+378 t^4+3653 t^6+27 t^5 (y^{-1}+y) + \ldots
\end{equation}
The coefficient $27$ is the dimension of the irreducible representation of $E_6$ and the coefficients $378$ and $3653$ are sums of dimensions of irreducible representations of $E_6$.

There is another very interesting observation made by Spiridonov and Vartanov in \cite{Spiridonov:2012ww}. It turns out that all 't Hooft anomaly matching conditions for dual theories can be derived from $SL(3,\mathbb{Z})$--modular transformation properties of the kernels of dual superconformal indices. Unfortunately,  a clear understanding of this relation is not known yet.

\section{Mirror duality via basic hypergeometric integrals}

In this section we will discuss mirror symmetry in three dimensions as an example of a three-dimensional supersymmetric duality. Mirror symmetry means that the infrared limit of one $3d$ ${\mathcal N}=2$ (or ${\mathcal N}=4$) theory is the same as the infrared limit of another ${\mathcal N}=2$ supersymmetric gauge theory. As in the Seiberg duality, the mirror dual theories have the same flavor symmetries, however different gauge groups. One of the novelties that appears in three dimensions, compared to four, is that the superconformal index includes sum over monopole charges. 

Let us consider the following two theories which are dual under the mirror symmetry \cite{Intriligator:1996ex, deBoer:1997ka, Aharony:1997bx}:
\begin{itemize}

\item \textbf{Theory A}: three-dimensional ${\mathcal N}=2$ supersymmetric field theory with $U(1)$ gauge group and one flavor. The superconformal index of this theory has the following form \cite{Imamura:2011su,Krattenthaler:2011da,Gahramanov:2013rda}
\begin{equation}  \label{mirA}
\sum_{m \in Z} q^{|m|/3} \int_{\mathbb{T}} \frac{dz}{2\pi i z} \frac{(q^{5/6+|m|/2}z;q)_{\infty}(q^{5/6+|m|/2}z^{-1};q)_{\infty}}{(q^{1/6+|m|/2}z;q)_{\infty}(q^{1/6+|m|/2}z^{-1};q)_{\infty}}
\end{equation}
where the sum is over monopole charges $m$.

\item \textbf{Theory B}: the free Wess--Zumino theory  with three chiral multiplets, without gauge degrees of freedom. The index of the theory is given by the simpler expression
\begin{equation} 
\left( \frac{(q^{2/3};q)_{\infty}}{(q^{1/3};q)_{\infty}} \right)^3
\end{equation}

\end{itemize}

The duality of the theories leads to the following integral identity for the basic hypergeometric integrals
\begin{equation}
\sum_{m \in Z} q^{|m|/3} \int_{\mathbb{T}} \frac{dz}{2\pi i z} \frac{(q^{5/6+|m|/2}z;q)_{\infty}(q^{5/6+|m|/2}z^{-1};q)_{\infty}}{(q^{1/6+|m|/2}z;q)_{\infty}(q^{1/6+|m|/2}z^{-1};q)_{\infty}} \ = \ \left( \frac{(q^{2/3};q)_{\infty}}{(q^{1/3};q)_{\infty}} \right)^3
\end{equation}
We refer the reader to the work \cite{Krattenthaler:2011da} for the details and the mathematical proof of this identity. One can obtain more complicated identities for basic hypergeometric integrals by considering other three-dimensional dualities, see, e.g. \cite{Krattenthaler:2011da,Gahramanov:2013rda,Gahramanov:2014ona, Gahramanov:2016wxi,GahSpir}. These identities are interesting for many reasons. They are related to partition functions for non-supersymmetric Chern-Simons theories, knot invariants, integrability, etc.

\section{Reduction and hyperbolic hypergeometric integrals}

In this section we will sketch of the reduction of the four-dimensional superconformal index to three-dimensional partition function \cite{Dolan:2011rp,Imamura:2011uw,Gadde:2011ia}.

Consider the reduction of four-dimensional ${\mathcal N}=1$ theory along $S^1$ (or $R$) with a twisted boundary condition. As a result we obtain a three-dimensional ${\mathcal N}=2$ theory on the squashed sphere
\begin{equation}
S_b^3\ := \ \{ \ b^2|z_1|^2+b^{-2}|z_2|^2 \ = \ 1 \;, (z_1,z_2)\in {\mathbb{C}}^2 \} \;,
\end{equation}
where $b$ is the squashing parameter. In the reduction procedure the superconformal index reduces to the partition function on squashed three-sphere. From the perspective of special functions the essential step in the reduction is scaling the fugacities as
\begin{equation} \label{scaling}
p=e^{2\pi i v \omega_1}, \;\; q=e^{2 \pi i v \omega_2}, \;\; z=e^{2 \pi i v u}, \;\; t_i=e^{2\pi i v \alpha_i} \; .
\end{equation}
and then taking the limit $v \rightarrow 0$ of the $4d$ superconformal index.  This procedure turns the elliptic gamma function into the hyperbolic gamma function
\begin{equation} \label{limit}
 \Gamma(e^{2 \pi \textup{i} v z};e^{2 \pi \textup{i} v \omega_1}, e^{2 \pi i v \omega_2}) \;\; \rightarrow \;\; e^{-\pi
i(2z-(\omega_1+\omega_2))/24 v\omega_1\omega_2} \; \gamma^{(2)}(z;\omega_1,\omega_2) 
\end{equation}
where $\gamma^{2}(z;\omega_1, \omega_2)$ denotes the hyperbolic gamma function\footnote{It is related to the quantum dilogarithm \cite{Spiridonov:2011hf}.} 
\begin{equation}
\gamma^{(2)}(u;\omega_1,\omega_2) = e^{-\frac{\textup{i} \pi}{2} 
B_{2,2}(u;\omega_1,\omega_2)} \frac{(e^{2 \pi i u/\omega_1}
e^{-2 \pi i
\omega_2/\omega_1}; e^{-2 \pi i
\omega_2/\omega_1})_\infty}{(e^{2 \pi i
u/\omega_2}; e^{2 \pi i \omega_1/\omega_2})_\infty},
\end{equation}
and $B_{2,2}(u;\omega_1,\omega_2)$ is the second order Bernoulli polynomial
\begin{equation}
 B_{2,2}(u;\omega_1,\omega_2) =
\frac{u^2}{\omega_1\omega_2} - \frac{u}{\omega_1} -
\frac{u}{\omega_2} + \frac{\omega_1}{6\omega_2} +
\frac{\omega_2}{6\omega_1} + \frac 12.
\end{equation}
An integral over the hyperbolic gamma functions is called a hyperbolic hypergeometric integral. Note that the hyperbolic hypergeometric integral is well-defined also for $|q|=1$ (where $q=e^{2 \pi i \omega_1/\omega_2}$). Recently non-trivial identities for this type of integrals have been studied in the mathematical literature, see, e.g. \cite{stokman2005hyperbolic,Bultthesis}.

As an example, let us consider the following duality \cite{Intriligator:1995ff}: 
\begin{itemize}

\item \textbf{Theory A:} four-dimensional ${\mathcal N}=1$ theory with the gauge group $SP(2 N)$ and flavor group $SU(2 N_f)$, with matter $X$ in the $({N(2N - 1)}/{2} - 1)$-dimensional traceless antisymmetric tensor representation of the gauge group and $2N_f$ chiral fields $Q$ in the fundamental representation of $SP(2N)$ and $SU(2 N_f)$. The field content with global charges is given in the following table 
\vspace{0.1cm}

\begin{center}
\begin{tabular}{|c|c|c|c|}
  \hline
    & $SP(2N)$ & $SU(2N_f)$ & $U(1)_R$ \\  \hline
  $Q$ & $f$ & $f$ & $2r=1-\frac{2(N+K)}{(K+1)N_f}$ \\
  $X$ & $T_A$ & 1 & $2s=\frac{2}{K+1}$ \\
\hline
\end{tabular}
\\ \vspace{0.2cm} 
\textbf Matter content of the \textbf{Theory A} with the $R$ charge assignment.
\end{center}

\vspace{0.1cm}

\item \textbf{Theory B:} the theory with the gauge group $SP(2\widetilde{N})$, where $\widetilde{N} \ = \ K(N_f-2)-N, \ \ \ K=1,2,\ldots$; with matter $Y$ in the antisymmetric traceless representation of the gauge group, $2N_f$ chiral superfields $q$ in the fundamental representation of the gauge group and anti-fundamental representation of the flavor group and gauge singlets $M_j$, where $j = 1, \ldots , K$. The field content with global charges is given in the following table 
\vspace{0.1cm}

\begin{center}
\begin{tabular}{|c|c|c|c|}
  \hline
   & $SP(2\widetilde{N})$ & $SU(2N_f)$ & $U(1)_R$ \\  \hline
  $q$ & $f$ & $\overline{f}$ & $2\widetilde{r}=1-\frac{2(\widetilde{N} +K)}{(K+1)N_f}$
  \\
  $Y$ & $T_A$ & 1 & $2s=\frac{2}{K+1}$ \\
  $M_j$  & 1 & $T_A$ & $2r_j=2\frac{K+j}{K+1}- 4 \frac{\widetilde{N} +K}{(K+1)N_f}$ \\
\hline
\end{tabular}
\\ \vspace{0.2cm} 
\textbf Matter content of the \textbf{Theory B} with the $R$ charge assignment.
\end{center}

\end{itemize}

Indeed, defining $U=(pq)^s=(pq)^{\frac{1}{K+1}}$, we find the following superconformal indices for these theories \cite{Spiridonov:2009za}
\begin{align}
    I_A &= \frac{(p;p)_{\infty}^{N} (q;q)_{\infty}^{N} }{2^N N!} \Gamma(U;p,q)^{N-1}  \\ \nonumber 
        & \quad \times  \int_{\mathbb{T}^N} \prod_{1 \leq i < j \leq N} \frac{\Gamma(U z_i^{\pm 1} z_j^{\pm
   1};p,q)}{\Gamma(z_i^{\pm 1} z_j^{\pm 1};p,q)}   \prod_{j=1}^{N} \frac{\prod_{i=1}^{2N_f}  \Gamma(s_i z_j^{\pm 1};p,q)} {\Gamma(z_j^{\pm 2};p,q)}\prod_{j=1}^{N} \frac{d z_j}{2 \pi \textup{i} z_j},\\
    I_B &=  \frac{(p;p)_{\infty}^{\widetilde{N}} (q;q)_{\infty}^{\widetilde{N}} }{2^{\widetilde{N}}
    \widetilde{N}!} \Gamma(U;p,q)^{\widetilde{N}-1} \prod_{l=1}^K \prod_{1 \leq i < j \leq 2N_f} \Gamma(U^{l-1} s_i s_j;p,q)
    \\ \nonumber
    & \quad  \times \int_{\mathbb{T}^{\widetilde{N}}}  \prod_{1 \leq i < j \leq \widetilde{N}} \frac{\Gamma(U z_i^{\pm 1} z_j^{\pm
    1};p,q)}{\Gamma(z_i^{\pm 1} z_j^{\pm 1};p,q)} \prod_{j=1}^{\widetilde{N}} \frac{\prod_{i=1}^{2N_f} \Gamma(U s_i^{-1} z_j^{ \pm 1};p,q)}
    {\Gamma(z_j^{\pm 2};p,q)}\prod_{j=1}^{\widetilde{N}}\frac{d z_j}{2 \pi \textup{i} z_j},
\end{align}
where the balancing condition reads $U^{2(N+K)}\prod_{i=1}^{2N_f}s_i= (pq)^{N_f}$. The duality leads to the identity $I_A=I_B$. By taking the limit (\ref{limit}) and canceling the identical ``infinite'' factors\footnote{We also integrated out one flavor.} we obtain the following non-trivial identity for the hyperbolic hypergeometric integrals \cite{Gahramanov:gka}:
\begin{align} \nonumber
\frac{1}{{\scriptstyle 2^N N!}}
\gamma({\scriptstyle \frac{\omega_1+\omega_2}{K+1}})^{N-1}
\int_{-\textup{i} \infty}^{\textup{i} \infty} \prod_{{\scriptscriptstyle 1 \leq i < j \leq N}}
\frac{\gamma({\scriptstyle\frac{\omega_1+\omega_2}{K+1} \pm u_i \pm
u_j})}{\gamma({\scriptstyle \pm u_i \pm
u_j})} \prod_{j=1}^{N} \frac{\prod_{i=1}^{2(N_f-1)}
\gamma({\scriptstyle \alpha_i \pm u_j})} {\gamma({\scriptstyle \pm
2 u_j})} \frac{d u_j}{{\scriptstyle \textup{i} \sqrt{\omega_1 \omega_2}}} \\ \nonumber
= \frac{1}{{\scriptstyle 2^{\widetilde{N}} \widetilde{N}!}}
\gamma({\scriptstyle \frac{\omega_1+\omega_2}{K+1}})^{\widetilde{N}-1} \prod_{l=1}^K \gamma\Big({\scriptstyle(\omega_1+\omega_2)\big( N_f-\frac{2N+2K-l+1}{K+1}\big) - \sum_{i=1}^{2(N_f-1)} \alpha_i} \Big) 
 \prod_{l=1}^K \prod_{1 \leq i < j \leq 2(N_f-1)} \gamma\big({\scriptstyle (l-1) \frac{\omega_1+\omega_2}{K+1} + \alpha_i + \alpha_j}\big) \\
 \quad \times \int_{-\textup{i}
\infty}^{\textup{i} \infty} \prod_{1 \leq i < j \leq \widetilde{N}}
\frac{\gamma({\scriptstyle \frac{\omega_1+\omega_2}{K+1} \pm u_i \pm
u_j})}{\gamma({\scriptstyle \pm u_i \pm
u_j})}  \prod_{j=1}^{\widetilde{N}}
\frac{\prod_{i=1}^{2(N_f-1)} \gamma({\scriptstyle \frac{\omega_1+\omega_2}{K+1}
- \alpha_i \pm u_j})} {\gamma({\scriptstyle \pm 2
u_j})} \prod_{j=1}^{\widetilde{N}} \frac{d
u_j}{\textup{i} {\scriptstyle\sqrt{\omega_1 \omega_2}}}.
\end{align}
where for convenience we used the shorthand $\gamma(u) := \gamma^{(2)}(u;\omega_1, \omega_2)$.

\section{Conclusions}

Recent progress in supersymmetric gauge theories have significant implications for mathematics. In these short notes, we presented relationships between partition functions for supersymmetric gauge theories on curved space-times and hypergeometric integrals. In particular, we focused on application of the hypergeometric integral identities to the verification of various supersymmetric dualities. This connection can open up many interesting directions for the future research.

\vspace{0.5cm}
\noindent \textbf{Acknowledgement.} I am very grateful to all my collaborators on the papers and projects that made possible this short review. The review is based on my talk given at the Geometry and Physics 2015 winter school in Srni, Czech Republic on January 17-24, 2015. I would like to thank the organizers, as well as the many participants with whom I had discussions. Some parts of the review were taken from my lecture notes on the course ``Elliptic hypergeometric functions and Physics'' given at the Undergraduate and Graduate Mathematics Summer School in Nesin Mathematics Village, Izmir, Turkey on September 15-21, 2014. I am grateful to the Nesin Mathematics Village for the hospitality. I especially thank Edoardo Vescovi for proof-reading the manuscript and suggesting valuable improvements for it.

\bibliographystyle{amsalpha}

\end{document}